\documentclass[twocolumn,dvipsnames]{aastex63}

\providecommand\aap{Astron.\ Astrophys.} % Astron and Astrophys
\providecommand\apjl{Astrophys.\ J.} % Astrophysical Journal, Letters
\providecommand\apjs{Astrophys.\ J.\ Suppl.} % Astrophysical Journal, Supplement
\received{August 25, 2021}
\revised{xxx}
\accepted{xxx}
\submitjournal{ApJ Letters}

\shorttitle{Physical Characterization of MBC (248370) 2005 QN$_{173}$}
\shortauthors{Hsieh et al.}
\begin{document}

\title{Physical Characterization of Main-Belt Comet (248370) 2005 QN$_{173}$}

\correspondingauthor{Henry H.\ Hsieh}
\email{hhsieh@psi.edu}

\author[0000-0001-7225-9271]{Henry H.\ Hsieh}
\affiliation{Planetary Science Institute, 1700 East Fort Lowell Rd., Suite 106, Tucson, AZ 85719, USA}
\affiliation{Institute of Astronomy and Astrophysics, Academia Sinica, P.O.\ Box 23-141, Taipei 10617, Taiwan}

\author[0000-0001-7335-1715]{Colin Orion\ Chandler}
\affiliation{Department of Astronomy and Planetary Science, Northern Arizona University, Flagstaff, AZ 86011, USA}

\author[0000-0002-7034-148X]{Larry Denneau}
\affiliation{Institute for Astronomy, University of Hawaii, 2680 Woodlawn Drive, Honolulu, HI 96822, USA}

\author[0000-0003-0250-9911]{Alan Fitzsimmons}
\affiliation{Astrophysics Research Centre, School of Physics and Astronomy, Queen's University Belfast, Belfast BT7 1NN, UK}

\author[0000-0002-9986-3898]{Nicolas Erasmus}
\affiliation{South African Astronomical Observatory, Cape Town, 7925, South Africa}

%\author[0000-0002-7332-2479]{Masateru Ishiguro}
%\affiliation{Department of Physics and Astronomy, Seoul National University, Gwanak, Seoul 151-742, Korea}

\author[0000-0002-6702-7676]{Michael S.\ P.\ Kelley}
\affiliation{Department of Astronomy, University of Maryland, 1113 Physical Sciences Complex, Building 415, College Park, MD 20742, USA}

\author[0000-0003-2781-6897]{Matthew M.\ Knight}
\affiliation{Department of Physics, U.S. Naval Academy, 572C Holloway Rd., Annapolis, MD, 21402, USA}
\affiliation{Department of Astronomy, University of Maryland, 1113 Physical Sciences Complex, Building 415, College Park, MD 20742, USA}

%\author[0000-0001-6765-6336]{Nicholas A.\ Moskovitz}
%\affiliation{Lowell Observatory, 1400 W.\ Mars Hill Rd, Flagstaff, AZ 86011, USA}

\author[0000-0002-3818-7769]{Tim A.\ Lister}
\affiliation{Las Cumbres Observatory, 6740 Cortona Drive Suite 102, Goleta, CA 93117, USA}

\author[0000-0002-5736-1857]{Jana Pittichov\'a}
\affiliation{Jet Propulsion Laboratory, California Institute of Technology, 4800 Oak Grove Dr, Pasadena, CA 91109, USA}

\author[0000-0003-3145-8682]{Scott S.\ Sheppard}
\affiliation{Earth and Planets Laboratory, Carnegie Institution for Science, 5241 Broad Branch Road NW, Washington, DC 20015, USA}

\author[0000-0002-1506-4248]{Audrey Thirouin}
\affiliation{Lowell Observatory, 1400 W.\ Mars Hill Rd, Flagstaff, AZ 86001, USA}

\author[0000-0001-9859-0894]{Chadwick A.\ Trujillo}
\affiliation{Department of Astronomy and Planetary Science, Northern Arizona University, Flagstaff, AZ 86011, USA}

\author[0000-0002-8658-5534]{Helen Usher}
\affiliation{The Open University, Walton Hall, Milton Keynes, MK7 6AA, UK}

\author[0000-0001-5749-1507]{Edward Gomez}
\affiliation{Las Cumbres Observatory, School of Physics and Astronomy, Cardiff University, Queens Buildings, The Parade, Cardiff CF24 3AA, UK}
%\ead{egomez@lco.global}

\author{Joey Chatelain}
\affiliation{Las Cumbres Observatory, 6740 Cortona Drive Suite 102, Goleta, CA 93117, USA}
%\ead{jchatelain@lco.global}

\author[0000-0002-4439-1539]{Sarah Greenstreet}
\affiliation{Asteroid Institute, 20 Sunnyside Ave, Suite 427, Mill Valley, CA 94941, USA}
\affiliation{Department of Astronomy and the DIRAC Institute, University of Washington, 3910 15th Ave NE, Seattle, WA 98195, USA}
%\ead{sarah@b612foundation.org}

\author{Tony Angel}
\affiliation{Harlingten Observatory, Observatorio Sierra Contraviesa, Cortijo El Cerezo, Torvizcon 18430, Granada, Spain}

\author{Richard Miles}
\affiliation{British Astronomical Association, UK}

\author[0000-0002-1977-065X]{Paul Roche}
\affiliation{Faulkes Telescope Project, School of Physics and Astronomy, Cardiff University, Cardiff, CF24 3AA, UK}

\author{Ben Wooding}
\affiliation{St Mary's Catholic Primary School, Llangewydd Road, Bridgend, Wales, CF31 4JW, UK}

%\author[0000-0002-4838-7676]{Quanzhi Ye}
%\affiliation{Department of Astronomy, University of Maryland, 1113 Physical Sciences Complex, Building 415, College Park, MD 20742, USA}

%\author{August Muench}
%\affiliation{American Astronomical Society \\
%1667 K Street NW, Suite 800 \\
%Washington, DC 20006, USA}

%\collaboration{1}{(AAS Journals Data Scientists collaboration)}

%\author{Butler Burton}
%\affiliation{Leiden University}
%\affiliation{AAS Journals Associate Editor-in-Chief}
%\nocollaboration{1}

%\author{Amy Hendrickson}
%\altaffiliation{AASTeX v6+ programmer}
%\affiliation{TeXnology Inc.}

%\collaboration{1}{(LaTeX collaboration)}

%\author{Julie Steffen}
%\affiliation{AAS Director of Publishing}
%\affiliation{American Astronomical Society \\
%1667 K Street NW, Suite 800 \\
%Washington, DC 20006, USA}

%% Mark off the abstract in the ``abstract'' environment. 
\begin{abstract}
We report results from new and archival observations of the newly discovered active asteroid (248370) 2005 QN$_{137}$, which has been determined to be a likely main-belt comet based on a subsequent discovery that it is recurrently active near perihelion. From archival data analysis, we estimate \mbox{$g'$-}, \mbox{$r'$-}, \mbox{$i'$-}, and \mbox{$z'$-}band absolute magnitudes for the nucleus of $H_g=16.62\pm0.13$, $H_r=16.12\pm0.10$, $H_i=16.05\pm0.11$, and $H_z=15.93\pm0.08$, corresponding to nucleus colors of $g'-r'=0.50\pm0.16$, $r'-i'=0.07\pm0.15$, and $i'-z'=0.12\pm0.14$, an equivalent $V$-band absolute magnitude of $H_V=16.32\pm0.08$, and a nucleus radius of $r_n=1.6\pm0.2$~km (using a $V$-band albedo of $p_V = 0.054\pm0.012$).  Meanwhile, we find mean near-nucleus coma colors when 248370 was active of $g'-r'=0.47\pm0.03$, $r'-i'=0.10\pm0.04$, and $i'-z'=0.05\pm0.05$, and similar mean dust tail colors, suggesting that no significant gas coma is present.  We find approximate ratios between the scattering cross-sections of near-nucleus dust (within 5000~km of the nucleus) and the nucleus of $A_d/A_n=0.7\pm0.3$ on 2016 July 22, and $1.8<A_d/A_n<2.9$ in 2021 July and August. During the 2021 observation period, the coma declined in intrinsic brightness by $\sim0.35$~mag (or $\sim$25\%) in 37 days, while the surface brightness of the dust tail remained effectively constant over the same period.  Constraints derived from the sunward extent of the coma suggest that terminal velocities of ejected dust grains are extremely slow ($\sim1~{\rm m~s}^{-1}$ for $1~\mu{\rm m}$ particles), indicating that the observed dust emission may have been aided by rapid rotation of the nucleus lowering the effective escape velocity.
\end{abstract}

%% Keywords should appear after the \end{abstract} command. 
%% See the online documentation for the full list of available subject
%% keywords and the rules for their use.

%% From the front matter, we move on to the body of the paper.
%% Sections are demarcated by \section and \subsection, respectively.
%% Observe the use of the LaTeX \label
%% command after the \subsection to give a symbolic KEY to the
%% subsection for cross-referencing in a \ref command.
%% You can use LaTeX's \ref and \label commands to keep track of
%% cross-references to sections, equations, tables, and figures.
%% That way, if you change the order of any elements, LaTeX will
%% automatically renumber them.
%%
%% We recommend that authors also use the natbib \citep
%% and \citet commands to identify citations.  The citations are
%% tied to the reference list via symbolic KEYs. The KEY corresponds
%% to the KEY in the \bibitem in the reference list below. 

\section{Introduction} \label{section:intro}

\begin{figure*}[htbp]
\centerline{\includegraphics[width=6.5in]{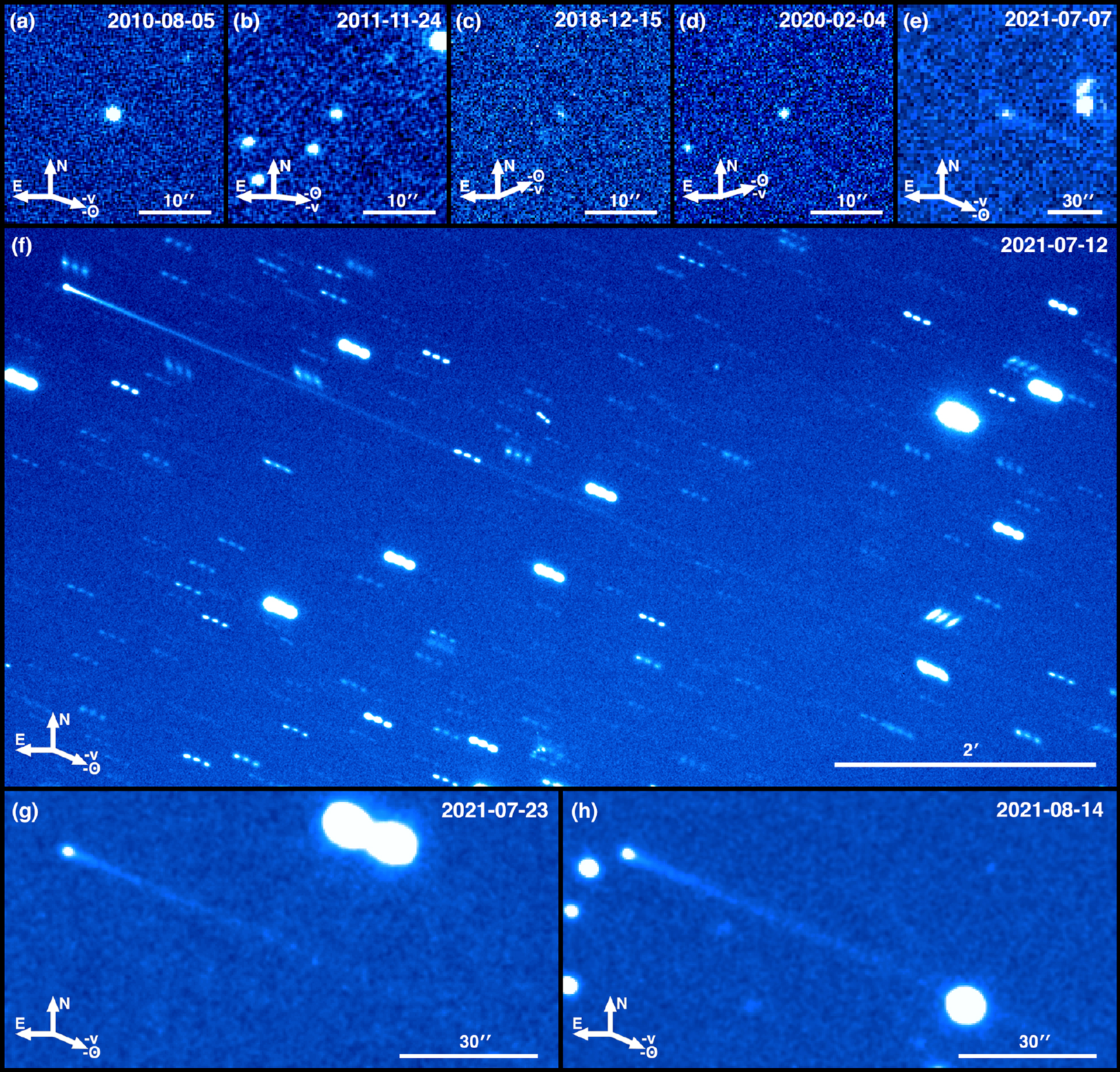}}
\caption{\small Single or composite images of 248370 for dates indicated in each panel (see Tables~\ref{table:obs_2005qn173_active} and \ref{table:obs_2005qn173_inactive} for observation details).  All images are in $r'$-band except for (e) which was obtained using the ATLAS survey's ``cyan'' filter (bandpass from 420~nm to 650~nm).  Scale bars indicate the size of each panel.  North (N), East (E), the antisolar direction ($-\odot$), and the negative heliocentric velocity direction ($-v$) are indicated in each panel.  The object is located at the center of panels (a) through (e), while in panels (f) through (h), the object's nucleus is located in the upper left corner with the tail extending down and to the right, where the latter set of images have been Gaussian-smoothed to enhance the visibility of low surface brightness features.
}
\label{figure:images}
\end{figure*}

Asteroid (248370) 2005 QN$_{173}$ (hereafter, 248370) was discovered to be active on UT 2021 July 7 in data comprising 120s of total exposure time (Figure~\ref{figure:images}e) obtained by the Asteroid Terrestrial-Impact Last Alert System \citep[ATLAS;][]{tonry2018_atlas} survey telescope \citep{fitzsimmons2021_2005QN173}.  On that date, the object was at a heliocentric distance of $r_h=2.391$~au and true anomaly of $\nu=16.0^{\circ}$, having most recently reached perihelion on UT 2021 May 14. As reported in the discovery announcement, 248370 exhibited a thin, straight dust tail $7\farcm6$ in length at a position angle of 245$^{\circ}$ East of North in confirmation observations obtained by Lowell Observatory's 4.3~m Lowell Discovery Telescope (LDT). Zwicky Transient Facility (ZTF) observations %\citep[ZTF;][]{smith2014_ztf} 
show the presence of the tail as early as UT 2021 June 11 \citep{kelley2021_2005QN173}.

As of 2021 August 1, 248370 has a semimajor axis of $a=3.067$~au, 
%perihelion distance of $q=2.374$~au, 
eccentricity of $e=0.226$, and inclination of $i=0.067^{\circ}$, according to the JPL Small-Body Database\footnote{\url{https://ssd.jpl.nasa.gov/sbdb.cgi}}, placing it unambiguously in the outer main asteroid belt.  248370's active nature and asteroidal orbit places it among the class of objects known as active asteroids, which exhibit comet-like mass loss yet have dynamically asteroidal orbits \citep{jewitt2015_actvasts_ast4}.  Active asteroids include main-belt comets \citep{hsieh2006_mbcs}, for which sublimation of volatile ice is the most likely activity driver, and disrupted asteroids, for which activity is due to other processes such as impacts or rotational destabilization \citep[e.g.,][]{hsieh2012_scheila}.  

248370 has been previously measured to have a diameter of 3.6$\pm$0.2~km and visible geometric albedo of 0.054$\pm$0.012, using $H_V=16.00$ for the V-band absolute magnitude and $G=0.15$ \citep{mainzer2019_neowisealbedos_v2}.  As of July 2021, there are no published rotational lightcurve data available for the object in the Asteroid Lightcurve Photometry Database\footnote{\url{https://minplanobs.org/alcdef/index.php}},
nor in the Asteroid Lightcurve Data Base archived by the NASA Planetary Data System (PDS).  Similarly, no taxonomic classification for 248370 is available in current PDS catalogs.
%nor in the Asteroid Lightcurve Data Base  \citep[LCDB;][]{warner2019_lcdb_v3} archived by the NASA Planetary Data System (PDS).  Similarly, no taxonomic classification is available in current PDS catalogs \citep[e.g.,][]{demeo2009_asteroidtaxonomy_pdss,neese2010_taxonomy,hasselmann2011_taxonomy}.
%According to the Asteroids Dynamic Site\footnote{\tt https://newton.spacedys.com/astdys/} (AstDyS-2), the object has no known family associations, and has a Lyapunov time of $t_{ly}=4.1$~kyr
%3-2-1

Following the discovery of 248370's activity in 2021, \citet{chandler2021_2005qn173_recurrentactivity} reported the discovery of activity in archival data from the Dark Energy Camera \citep[DECam;][]{flaugher2015_decam} on the 4~m Victor M.\ Blanco Telescope (hereafter, Blanco) at Cerro Tololo Interamerican Observatory (CTIO) obtained on UT 2016 July 22, when the object was at a true anomaly of $\nu=56.5^{\circ}$, having then most recently passed perihelion on UT 2016 January 3.  This discovery of two separate active apparitions of 248370, both near perihelion, is considered a strong indication that sublimation is responsible for the observed activity \citep[e.g.,][Chandler et al., submitted]{hsieh2012_scheila}.

\section{Observations}\label{section:observations}

\setlength{\tabcolsep}{3.6pt}
\setlength{\extrarowheight}{0em}
\begin{table*}[htb!]
\caption{(248370) 2005 QN$_{173}$ Activity Observations}
\centering
\smallskip
\footnotesize
\begin{tabular}{lcrccrccrcccrc}
\hline\hline
\multicolumn{1}{c}{UT Date}
 & \multicolumn{1}{c}{Telescope}
 & \multicolumn{1}{c}{$t$$^a$}
 & \multicolumn{1}{c}{Filter}
 & \multicolumn{1}{c}{$\theta_s$$^b$}
 & \multicolumn{1}{c}{$\nu$$^c$}
 & \multicolumn{1}{c}{$r_h$$^d$}
 & \multicolumn{1}{c}{$\Delta$$^e$}
 & \multicolumn{1}{c}{$\alpha$$^f$}
 & \multicolumn{1}{c}{$m(r_h,\Delta,\alpha)$$^g$}
 & \multicolumn{1}{c}{$m(1,1,0)$$^h$}
 & \multicolumn{1}{c}{$A_d/A_n$$^i$}
 & \multicolumn{1}{c}{$Af\rho$$^j$}
 & \multicolumn{1}{c}{$\Sigma_t$$^k$}
 \\
\hline
%2011 Jul 18 & \multicolumn{4}{l}{Perihelion..........} & 0.0 & 2.438 & 2.292 & 24.6 & --- \\
2021 Jul  7 & ATLAS  &  120 & cyan & 5.2 & 16.0 & 2.391 & 1.930 & 24.4 & --- & --- & --- & \multicolumn{1}{c}{---} & --- \\
\hline
2021 Jul  9 & Palomar  &  800 & $g'$ & 2.1 & 16.6 & 2.392 & 1.909 & 24.2 & 19.57$\pm$0.01 & 15.14$\pm$0.01 & 2.9$\pm$0.5 & 15.1$\pm$0.7 & 24.24$\pm$0.20 \\
2021 Jul  9 & FTN      &  480 & $g'$ & 1.5 & 16.6 & 2.392 & 1.909 & 24.2 & 19.71$\pm$0.01 & 15.28$\pm$0.01 & 2.4$\pm$0.5 & 13.6$\pm$0.8 & 23.87$\pm$0.20 \\
2021 Jul 12 & Palomar  &  300 & $g'$ & 1.4 & 17.5 & 2.394 & 1.877 & 23.8 & 19.52$\pm$0.02 & 15.14$\pm$0.02 & 2.9$\pm$0.5 & 16.4$\pm$0.8 & 23.66$\pm$0.20 \\
2021 Jul 13 & FTN      &  480 & $g'$ & 2.4 & 17.8 & 2.395 & 1.867 & 23.7 & 19.52$\pm$0.02 & 15.15$\pm$0.02 & 2.9$\pm$0.5 & 14.5$\pm$0.8 & 24.01$\pm$0.20 \\
2021 Jul 15 & FTN      &  480 & $g'$ & 1.5 & 18.5 & 2.396 & 1.842 & 23.4 & 19.55$\pm$0.02 & 15.22$\pm$0.02 & 2.6$\pm$0.5 & 14.6$\pm$0.8 & 23.77$\pm$0.20 \\
\hline
2021 Jul  8 & LDT      & 2400 & $r'$ & 1.8 & 16.3 & 2.391 & 1.920 & 24.3 & 19.14$\pm$0.01 & 14.69$\pm$0.01 & 2.7$\pm$0.4 & 23.2$\pm$0.9 & 23.45$\pm$0.20 \\
2021 Jul  9 & Palomar  & 1300 & $r'$ & 2.1 & 16.6 & 2.392 & 1.909 & 24.2 & 19.11$\pm$0.01 & 14.68$\pm$0.01 & 2.8$\pm$0.4 & 22.8$\pm$0.8 & 23.68$\pm$0.20 \\
2021 Jul  9 & FTN      &  480 & $r'$ & 1.4 & 16.6 & 2.392 & 1.909 & 24.2 & 19.19$\pm$0.01 & 14.76$\pm$0.01 & 2.5$\pm$0.4 & 22.2$\pm$1.0 & 23.34$\pm$0.20 \\
2021 Jul 10 & LCOGT 1m   &  480 & $r'$ & 1.8 & 16.9 & 2.393 & 1.898 & 24.1 & 19.17$\pm$0.02 & 14.76$\pm$0.02 & 2.5$\pm$0.4 & 21.4$\pm$1.0 & 23.42$\pm$0.20 \\
2021 Jul 12 & Palomar  &  900 & $r'$ & 1.2 & 17.5 & 2.394 & 1.877 & 23.8 & 19.12$\pm$0.01 & 14.74$\pm$0.01 & 2.6$\pm$0.4 & 23.2$\pm$0.9 & 23.17$\pm$0.20 \\
2021 Jul 13 & FTN      &  480 & $r'$ & 2.2 & 17.8 & 2.395 & 1.867 & 23.7 & 19.05$\pm$0.01 & 14.68$\pm$0.01 & 2.8$\pm$0.4 & 22.6$\pm$0.9 & 23.55$\pm$0.20 \\
2021 Jul 14 & LCOGT 1m   & 1130 & $r'$ & 2.1 & 18.1 & 2.395 & 1.856 & 23.6 & 19.11$\pm$0.02 & 14.76$\pm$0.02 & 2.5$\pm$0.4 & 20.8$\pm$1.0 & 23.66$\pm$0.20 \\
2021 Jul 15 & LCOGT 1m   & 1130 & $r'$ & 1.7 & 18.5 & 2.396 & 1.842 & 23.4 & 19.05$\pm$0.01 & 14.72$\pm$0.01 & 2.6$\pm$0.4 & 22.8$\pm$0.9 & 23.41$\pm$0.20 \\
2021 Jul 15 & FTN      &  480 & $r'$ & 1.4 & 18.5 & 2.396 & 1.842 & 23.4 & 19.13$\pm$0.01 & 14.80$\pm$0.01 & 2.4$\pm$0.3 & 21.2$\pm$1.0 & 23.26$\pm$0.20 \\
2021 Jul 18 & LCOGT 1m   &  283 & $r'$ & 1.7 & 19.3 & 2.398 & 1.815 & 23.0 & 19.03$\pm$0.06 & 14.75$\pm$0.06 & 2.5$\pm$0.4 & 22.1$\pm$1.9 & 23.40$\pm$0.20 \\
2021 Jul 19 & LCOGT 1m   & 1130 & $r'$ & 1.9 & 19.5 & 2.399 & 1.805 & 22.9 & 19.06$\pm$0.02 & 14.78$\pm$0.02 & 2.4$\pm$0.4 & 20.7$\pm$1.0 & 23.45$\pm$0.20 \\
2021 Jul 21 & LCOGT 1m   & 1130 & $r'$ & 2.2 & 20.1 & 2.401 & 1.785 & 22.5 & 19.07$\pm$0.03 & 14.83$\pm$0.03 & 2.3$\pm$0.3 & 18.8$\pm$1.1 & 23.75$\pm$0.20 \\
2021 Jul 23 & LCOGT 1m   & 1853 & $r'$ & 1.9 & 20.8 & 2.402 & 1.762 & 22.1 & 19.05$\pm$0.03 & 14.85$\pm$0.03 & 2.2$\pm$0.3 & 19.0$\pm$1.1 & 23.52$\pm$0.20 \\
2021 Aug  5 & LCOGT 1m   &  600 & $r'$ & 2.1 & 24.5 & 2.414 & 1.648 & 19.2 & 18.86$\pm$0.02 & 14.89$\pm$0.02 & 2.1$\pm$0.3 & 17.9$\pm$1.0 & 23.55$\pm$0.20 \\
2021 Aug  7 & LCOGT 1m   &  600 & $r'$ & 2.0 & 25.1 & 2.415 & 1.632 & 18.7 & 18.80$\pm$0.01 & 14.87$\pm$0.01 & 2.2$\pm$0.3 & 18.7$\pm$0.9 & 23.50$\pm$0.20 \\
%2021 Aug 10 & LCOGT 1m   &  600 & $r'$ & 2.1 & 25.9 & 2.418 & 1.608 & 17.8 & 18.58$\pm$0.01 & 14.71$\pm$0.01 & 2.7$\pm$0.4 & 22.9$\pm$1.0 & 23.47$\pm$0.10 \\ % on field star
2021 Aug 14 & LCOGT 1m   &  600 & $r'$ & 2.0 & 27.2 & 2.423 & 1.577 & 16.4 & 18.81$\pm$0.01 & 15.02$\pm$0.02 & 1.8$\pm$0.3 & 15.3$\pm$1.0 & 23.45$\pm$0.20 \\
\hline
2021 Jul  9 & Palomar  &  900 & $i'$ & 2.1 & 16.6 & 2.392 & 1.909 & 24.2 & 18.98$\pm$0.01 & 14.55$\pm$0.01 & 2.8$\pm$0.2 & 25.7$\pm$0.5 & 23.60$\pm$0.20 \\
2021 Jul  9 & FTN      &  480 & $i'$ & 1.3 & 16.6 & 2.392 & 1.909 & 24.2 & 19.16$\pm$0.02 & 14.73$\pm$0.02 & 2.2$\pm$0.2 & 22.2$\pm$0.9 & 23.07$\pm$0.20 \\
2021 Jul 12 & Palomar  &  600 & $i'$ & 1.2 & 17.5 & 2.394 & 1.877 & 23.8 & 19.03$\pm$0.01 & 14.65$\pm$0.01 & 2.4$\pm$0.2 & 24.8$\pm$0.6 & 22.94$\pm$0.20 \\
2021 Jul 13 & FTN      &  480 & $i'$ & 2.2 & 17.8 & 2.395 & 1.867 & 23.7 & 18.92$\pm$0.02 & 14.55$\pm$0.02 & 2.8$\pm$0.2 & 25.5$\pm$0.9 & 23.40$\pm$0.20 \\
2021 Jul 15 & FTN      &  480 & $i'$ & 1.4 & 18.5 & 2.396 & 1.842 & 23.4 & 18.95$\pm$0.02 & 14.62$\pm$0.02 & 2.5$\pm$0.2 & 25.5$\pm$0.9 & 23.07$\pm$0.20 \\
\hline
2016 Jul 22 & Blanco   &   89 & $z'$ & 1.1 & 56.5 & 2.591 & 2.571 & 22.7 & 20.51$\pm$0.10 & 15.31$\pm$0.10 & 0.7$\pm$0.3 &  7.5$\pm$2.4 & 25.30$\pm$0.50 \\
2021 Jul  9     & FTN      &  480 & $z'$ & 1.2 & 16.6 & 2.392 & 1.909 & 24.2 & 19.05$\pm$0.04 & 14.62$\pm$0.04 & 2.2$\pm$0.5 & 24.5$\pm$2.1 & 23.08$\pm$0.30 \\
2021 Jul 13     & FTN      &  480 & $z'$ & 2.1 & 17.8 & 2.395 & 1.867 & 23.7 & 18.84$\pm$0.03 & 14.48$\pm$0.03 & 2.6$\pm$0.5 & 27.2$\pm$1.9 & 23.40$\pm$0.30 \\
2021 Jul 15     & FTN      &  480 & $z'$ & 1.3 & 18.5 & 2.396 & 1.842 & 23.4 & 18.92$\pm$0.03 & 14.59$\pm$0.03 & 2.2$\pm$0.5 & 25.5$\pm$1.9 & 23.06$\pm$0.30 \\
%2021 Jul 24 & FTN     & 4 &  480 & $g'$ & 21.0 & 2.403 & 1.756 & 22.0 & 19.47$\pm$0.05 & 15.28$\pm$0.05 &  \\  % 60 deg from 99.9% full moon
%2021 Jul 24 & FTN     & 4 & 480 & $r'$ &  21.0 & 2.403 & 1.756 & 22.0 & 18.95$\pm$0.03 & 14.76$\pm$0.03 &  \\  % 60 deg from 99.9% full moon
%2021 Jul 24 & FTN     & 4 & 480 & $i'$ &  21.0 & 2.403 & 1.756 & 22.0 & 18.96$\pm$0.04 & 14.77$\pm$0.04 &  \\  % 60 deg from 99.9% full moon
%2021 Jul 24 & FTN     & 4 & 480 & $z'$ &  21.0 & 2.403 & 1.756 & 22.0 & 18.74$\pm$0.04 & 14.55$\pm$0.04 &  \\  % 60 deg from 99.9% full moon
%\hline
%\hline
\hline
\hline
\multicolumn{14}{l}{$^a$ Total integration time, in seconds.} \\
\multicolumn{14}{l}{$^b$ FWHM seeing, in arcseconds.} \\
\multicolumn{14}{l}{$^c$ True anomaly, in degrees.} \\
\multicolumn{14}{l}{$^d$ Heliocentric distance, in au.} \\
\multicolumn{14}{l}{$^e$ Geocentric distance, in au.} \\
\multicolumn{14}{l}{$^f$ Solar phase angle (Sun-object-Earth), in degrees.} \\
\multicolumn{14}{l}{$^g$ Mean apparent coma magnitude (measured using 5000~km-radius photometry apertures).} \\
\multicolumn{14}{l}{$^h$ Computed absolute magnitude corresponding to measured apparent magnitude assuming a $H,G$ phase function where $G=0.15$.} \\
\multicolumn{14}{l}{$^i$ Inferred ratio of effective scattering cross-sectional areas of coma dust (within 5000~km-radius photometry apertures) and the nucleus,} \\
\multicolumn{14}{l}{$~~~$  where listed uncertainties only reflect photometric measurement uncertainty and not potential nucleus rotational variability.} \\
\multicolumn{14}{l}{$^j$ $A(\alpha=0^{\circ})f\rho$ values, computed using Equation~\ref{equation:afrho} and 5000~km photometry apertures, in cm.} \\
\multicolumn{14}{l}{$^k$ Dust tail surface brightness in mag~arcsec$^{-2}$, as measured in a 1500~km $\times$ 10000~km rectangular aperture, as described in the text.} \\
\end{tabular}
\label{table:obs_2005qn173_active}
\end{table*}

\setlength{\tabcolsep}{4.5pt}
\setlength{\extrarowheight}{0em}
\begin{table*}[htb!]
\caption{(248370) 2005 QN$_{173}$ Nucleus Observations}
\centering
\smallskip
\footnotesize
\begin{tabular}{lccrcrccrcc}
\hline\hline
\multicolumn{1}{c}{UT Date}
 & \multicolumn{1}{c}{Tel.$^a$}
 & \multicolumn{1}{c}{$N$$^b$}
 & \multicolumn{1}{c}{$t$$^c$}
 & \multicolumn{1}{c}{Filter}
 & \multicolumn{1}{c}{$\nu$$^d$}
 & \multicolumn{1}{c}{$r_h$$^e$}
 & \multicolumn{1}{c}{$\Delta$$^f$}
 & \multicolumn{1}{c}{$\alpha$$^g$}
 & \multicolumn{1}{c}{$m(r_h,\Delta,\alpha)$$^h$}
 & \multicolumn{1}{c}{$m(1,1,0)$$^i$}
 \\
\hline
%2011 Jul 18 & \multicolumn{4}{l}{Perihelion..........} & 0.0 & 2.438 & 2.292 & 24.6 & --- \\
2010 Aug  6 & PS1      & 1 &   43 & $g'$ & 354.7 & 2.389 & 1.377 &  2.0 & 19.17$\pm$0.03 & 16.35$\pm$0.03 \\
2010 Sep  6 & PS1      & 2 &   86 & $g'$ &   3.8 & 2.388 & 1.463 & 12.2 & 20.08$\pm$0.05 & 16.63$\pm$0.05 \\
%2010 Sep 12 & PS1      & 2 &   86 & $g'$ &   5.6 & 2.390 & 1.504 & 14.5 &  &  \\  % something wrong; photometry too faint
2011 Nov 24 & PS1      & 2 &   86 & $g'$ & 109.4 & 3.150 & 2.180 &  4.1 & 21.20$\pm$0.13 & 16.64$\pm$0.13 \\
%2011 Nov 25 & PS1      & 2 &   86 & $g'$ & 109.5 & 3.152 & 2.179 &  3.7 &  &  \\  % object not visible
2011 Dec  1 & PS1      & 2 &   86 & $g'$ & 110.6 & 3.164 & 2.180 &  1.4 & 20.98$\pm$0.10 & 16.60$\pm$0.10 \\
%2013 Jan 16 & PS1      & 3 &  129 & $g'$ & 167.7 & 3.728 & 2.775 &  4.4 &  &  \\  % object not visible
%2013 Mar 04 & PS1      & 2 &   86 & $g'$ & 173.4 & 3.746 & 2.922 &  9.6 &  &  \\  % object not visible
{\it Median$^j$} & ... & ... & ... & $g'$ & ... & ... & ... & ... & ... & 16.62$\pm$0.13 \\
\hline
2010 Aug  5 & PS1      & 1 &   40 & $r'$ & 354.4 & 2.389 & 1.378 &  2.5 & 18.81$\pm$0.03 & 15.95$\pm$0.03 \\
%2010 Sep 12 & PS1      & 2 &   80 & $r'$ &   5.6 & 2.390 & 1.504 & 14.5 &  &  \\  % something wrong; photometry too faint
%2011 Sep 07 & PS1      & 1 &   40 & $r'$ &  95.3 & 2.979 & 2.839 & 19.8 &  &  \\  % object not visible
%2011 Sep 10 & PS1      & 1 &   40 & $r'$ &  95.9 & 2.985 & 2.804 & 19.7 &  &  \\  % object not visible
2011 Nov 24 & PS1      & 2 &   80 & $r'$ & 109.4 & 3.150 & 2.180 &  4.0 & 20.74$\pm$0.10 & 16.19$\pm$0.10 \\
%2011 Nov 25 & PS1      & 2 &   80 & $r'$ & 109.6 & 3.152 & 2.179 &  3.7 &  &  \\  % object not visible
%2011 Dec 01 & PS1      & 1 &   40 & $r'$ & 110.6 & 3.164 & 2.180 &  1.4 &  &  \\  % object not visible
%2013 Jan 23 & PS1      & 2 &   80 & $r'$ & 168.6 & 3.731 & 2.755 &  2.2 &  &  \\  % object not visible
2018 Dec 15 & Blanco & 1 &   45 & $r'$ & 191.7 & 3.733 & 3.508 & 15.2 & 22.63$\pm$0.31 & 16.20$\pm$0.31 \\  % only one image
2020 Feb  4 & Blanco & 2 &   81 & $r'$ & 248.9 & 3.165 & 3.059 & 18.1 & 21.91$\pm$0.11 & 16.04$\pm$0.11 \\  % only one image in frame
%2020 Jun 17 & ZTF$^k$ & ... & ... & $r'$ & 273.7 & 2.867 & 2.072 & 14.9 & 20.33$\pm$0.20 & 15.63$\pm$0.20 \\
%2021 Jun 11 & ZTF$^k$ & ... & ... & $r'$ &   8.3 & 2.378 & 2.220 & 25.2 & 19.30$\pm$0.12 & 14.53$\pm$0.12 \\
%2021 Jun 17 & ZTF$^k$ & ... & ... & $r'$ &  10.1 & 2.380 & 2.152 & 25.3 & 19.34$\pm$0.11 & 14.63$\pm$0.11 \\
%2021 Jun 27 & ZTF$^k$ & ... & ... & $r'$ &  13.1 & 2.385 & 2.040 & 25.0 & 19.11$\pm$0.20 & 14.52$\pm$0.20 \\
%2021 Jul  7 & ZTF$^k$ & ... & ... & $r'$ &  16.0 & 2.391 & 1.930 & 24.4 & 19.12$\pm$0.09 & 14.66$\pm$0.09 \\
%2021 Jul 10 & ZTF$^k$ & ... & ... & $r'$ &  16.9 & 2.393 & 1.898 & 24.1 & 19.16$\pm$0.09 & 14.75$\pm$0.09 \\
{\it Median$^j$} & ... & ... & ... & $r'$ & ... & ... & ... & ... & ... & 16.12$\pm$0.10 \\
\hline
%2000 Dec 14 & INT      & 3 & 2157 & $i'$ &  98.0 & 3.034 & 2.172 & 10.6 &  &  \\  # unreduced data
2004 Jul  8 & CFHT     & 3 &  540 & $i'$ & 287.5 & 2.740 & 2.028 & 17.7 & 20.72$\pm$0.03 & 16.07$\pm$0.03 \\
2010 Aug  2 & PS1      & 1 &   45 & $i'$ & 353.5 & 2.390 & 1.383 &  3.9 & 19.01$\pm$0.05 & 16.05$\pm$0.05 \\
2010 Aug 31 & PS1      & 2 &   90 & $i'$ &   2.1 & 2.388 & 1.429 &  9.8 & 19.23$\pm$0.04 & 15.93$\pm$0.04 \\
2011 Nov 30 & PS1      & 2 &   90 & $i'$ & 110.4 & 3.162 & 2.179 &  1.9 & 20.26$\pm$0.09 & 15.84$\pm$0.09 \\
2015 Aug 18 & SkyMapper & 1 & 100 & $i'$ & 320.3 & 2.483 & 1.888 & 21.8 & 20.72$\pm$0.26 & 16.31$\pm$0.26 \\
%2013 Jan 21 & PS1      & 2 &   90 & $i'$ & 168.3 & 3.731 & 2.759 &  2.8 &  &  \\  % object not visible
%2013 Jan 22 & PS1      & 2 &   90 & $i'$ & 168.5 & 3.731 & 2.757 &  2.5 &  &  \\  % object not visible
%2014 May 17 & PS1      & 6 &  270 & $i'$ & 228.9 & 3.415 & 2.813 & 15.1 &  &  \\  % object not visible
{\it Median$^j$} & ... & ... & ... & $i'$ & ... & ... & ... & ... & ... & 16.05$\pm$0.11 \\
\hline
%2009 Jun 14 & PS1      & 2 &   60 & $z'$ & 254.6 & 3.100 & 2.518 & 17.1 &  &  \\
%2009 Jun 15 & PS1      & 2 &   60 & $z'$ & 254.8 & 3.098 & 2.529 & 17.1 &  &  \\
2010 Jun 14 & PS1      & 2 &   60 & $z'$ & 339.3 & 2.416 & 1.728 & 21.1 & 20.07$\pm$0.13 & 15.93$\pm$0.13 \\
%2010 Jun 22 & PS1      & 2 &   60 & $z'$ & 341.6 & 2.410 & 1.644 & 19.3 &  &  \\
2010 Oct 30 & PS1      & 5 &  150 & $z'$ &  19.6 & 2.413 & 2.018 & 23.8 & 20.48$\pm$0.23 & 15.93$\pm$0.13 \\
%2012 Feb 05 & PS1      & 2 &   60 & $z'$ & 121.2 & 3.297 & 2.820 & 16.3 &  &  \\
%2012 Oct 31 & PS1      & 2 &   60 & $z'$ & 158.3 & 3.678 & 3.654 & 15.6 &  &  \\
%2012 Dec 15 & PS1      & 2 &   60 & $z'$ & 163.8 & 3.710 & 3.037 & 12.3 &  &  \\
%2013 Apr 24 & PS1      & 2 &   60 & $z'$ & 179.5 & 3.753 & 3.601 & 15.5 &  &  \\
%2014 Jan 04 & PS1      & 1 &   60 & $z'$ & 210.8 & 3.606 & 3.359 & 15.7 &  &  \\
%2016 Jul 22 & Blanco & 1 &   89 & $z'$ &  56.5 & 2.591 & 2.571 & 22.7 & 20.55$\pm$0.10 & 15.35$\pm$0.10 \\  % only one image / active
%2016 Jul 23 & Blanco & 1 &  137 & $z'$ &  56.8 & 2.593 & 2.560 & 22.7 &  &  \\  % target not in FOV
%2016 Jul 24 & Blanco & 1 &  176 & $z'$ &  57.0 & 2.595 & 2.549 & 22.8 &  &  \\  % target not in FOV
2020 Feb 10 & Blanco & 1 &  199 & $z'$ & 250.0 & 3.152 & 2.960 & 18.2 & 21.59$\pm$0.14 & 15.80$\pm$0.14 \\  % only one image
{\it Median$^j$} & ... & ... & ... & $z'$ & ... & ... & ... & ... & ... & 15.93$\pm$0.08 \\
\hline
\hline
\multicolumn{11}{l}{$^a$ Telescope used} \\
\multicolumn{11}{l}{$^b$ Number of exposures.} \\
\multicolumn{11}{l}{$^c$ Total integration time, in seconds.} \\
\multicolumn{11}{l}{$^d$ True anomaly, in degrees.} \\
\multicolumn{11}{l}{$^e$ Heliocentric distance, in au.} \\
\multicolumn{11}{l}{$^f$ Geocentric distance, in au.} \\
\multicolumn{11}{l}{$^g$ Solar phase angle (Sun-object-Earth), in degrees.} \\
\multicolumn{11}{l}{$^h$ Mean apparent magnitude in specified filter.} \\
\multicolumn{11}{l}{$^i$ Computed absolute magnitude corresponding to measured apparent magnitude assuming IAU $H,G$ } \\
\multicolumn{11}{l}{$~~~$ phase function behavior where $G=0.15$.} \\
\multicolumn{11}{l}{$^j$ Median values of computed absolute magnitudes, where standard deviations are used as uncertainties.} \\
%\multicolumn{11}{l}{$^k$ Photometry reported by \citet{kelley2021_2005QN173}.} \\
\end{tabular}
\label{table:obs_2005qn173_inactive}
\end{table*}

New observations of 248370 were obtained on several nights between UT 2021 July 8 and UT 2021 August 14 with LDT \citep{levine2012_ldt}, Palomar Observatory's 5~m Hale Telescope (hereafter, Palomar), the 2~m Faulkes Telescope North (FTN), and Las Cumbres Observatory (hereafter, LCOGT) 1~m telescopes \citep{brown2013_lcogt} at CTIO and the South African Astronomical Observatory (SAAO) (Table~\ref{table:obs_2005qn173_active}).
Observations were obtained using the LDT's Large Monolithic Imager \citep[LMI;][]{bida2014_dct}, Palomar's Wafer-Scale camera for Prime \citep[WaSP;][]{nikzad2017_detectors} wide field prime focus camera, FTN's Multicolor Simultaneous Camera for studying Atmospheres of Transiting exoplanets \citep[MuSCAT3;][]{narita2020_muscat3}, and LCOGT Sinistro cameras.  All observations were obtained using Sloan $g'$-, $r'$-, $i'$-, or $z'$-band filters, and non-sidereal tracking to follow the target's motion.

Multi-filter FTN data were obtained using the simultaneous $g'$-, $r'$-, $i'$-, and $z'$-band imaging capability of MuSCAT3.
Multi-filter Palomar observations were obtained by interspersing filters (i.e., using repeating $r'g'r'i'r'$ or $r'i'g'r'$ sequences) to enable the use of interpolation to approximate simultaneous multi-filter imaging for color computation (i.e., compensating for possible rotational variability in the nucleus brightness between our actual observations in different filters).

Bias subtraction, flat field correction, and cosmic ray removal were performed for LDT and Palomar data using Python 3 code utilizing the {\tt ccdproc} package in Astropy \citep{astropy2018_astropy} and {\tt L.A.Cosmic} code\footnote{Written for python by Maltes Tewes (\url{https://github.com/RyleighFitz/LACosmics})} \citep{vandokkum2001_lacosmic,vandokkum2012_lacosmic}.  FTN and LCOGT 1~m data were processed using standard LCOGT pipeline software \citep{mccully2018_banzai}.
%\citep[BANZAI;][]{mccully2018_banzai}.
%to perform the standard bias and dark subtraction, flat field correction and astrometric solution.

We also used the Canadian Astronomy Data Centre's Solar System Object Image Search tool\footnote{\url{http://www.cadc-ccda.hia-iha.nrc-cnrc.gc.ca/en/ssois/}} \citep{gwyn2012_ssois} and the NASA Planetary Data System (PDS) Small Bodies Node's Comet Asteroid Telescopic Catalog Hub (CATCH) tool\footnote{\url{https://catch.astro.umd.edu/}} to identify archival Sloan $g'$-, $r'$-, $i'$-, and $z'$-band observations of 248370 from 2004 to 2020 (Table~\ref{table:obs_2005qn173_inactive}) from the 1.8~m Panoramic Survey Telescope and Rapid Response System (Pan-STARRS1; hereafter PS1) survey telescope \citep{chambers2016_panstarrs,flewelling2020_ps1}, MegaCam \citep{boulade2003_megacam} on the 3.6~m Canada-France-Hawaii Telescope (CFHT), the 1.35~m SkyMapper survey telescope \citep{wolf2018_skymapper_dr1}, and Blanco. For the purposes of our analysis, PS1 $g_{P1}$, $r_{P1}$, $i_{P1}$, and $z_{P1}$ filters are considered functionally equivalent to their Sloan counterparts.  All archival data were pipeline-processed by their respective facilities.
%, and as such, we did not perform any additional basic image processing of any of these images. %\hhh{need telescope/instrument citations}.
%\msk{I think I have a SkyMapper detection in 2015.  Need to look into this some more... Seems to be a good detection of an inactive comet.}

The object was identified in archival images either from its non-sidereal motion when more than one image was available on a particular night, or otherwise from comparison with reference images obtained on other nights when the object was not in the field of view.

\setlength{\tabcolsep}{4.5pt}
\setlength{\extrarowheight}{0em}
\begin{table*}[htb!]
\caption{(248370) 2005 QN$_{173}$ Colors}
\centering
\smallskip
\footnotesize
\begin{tabular}{lccccccccr}
\hline\hline
 & & & \multicolumn{3}{c}{Coma} & & \multicolumn{3}{c}{Tail} \\
\multicolumn{1}{c}{UT Date}
 & \multicolumn{1}{c}{Telescope} &
 & \multicolumn{1}{c}{$g'-r'$}
 & \multicolumn{1}{c}{$r'-i'$}
 & \multicolumn{1}{c}{$i'-z'$} &
 & \multicolumn{1}{c}{$g'-r'$}
 & \multicolumn{1}{c}{$r'-i'$}
 & \multicolumn{1}{c}{$i'-z'$}
 \\
\hline
2021 Jul  9 & Palomar & & 0.47$\pm$0.01 & 0.13$\pm$0.01 & --- & & 0.56$\pm$0.30 & 0.08$\pm$0.30 & --- \\
% coma: g-r=0.471+/-0.009; r-i=0.134+/-0.007
% tail: g-r=0.470+/-0.234; r-i=0.025+/-0255
2021 Jul  9 & FTN & & 0.52$\pm$0.02 & 0.03$\pm$0.03 & 0.14$\pm$0.04 & & 0.53$\pm$0.30 & 0.27$\pm$0.30 & $-$0.02$\pm$0.40 \\
% coma: g-r=0.535+/-0.019; r-i=0.034+/-0.027; i-z=0.110+/-0.040
2021 Jul 12 & Palomar & & 0.42$\pm$0.02 & 0.08$\pm$0.01 & --- & & 0.49$\pm$0.30 & 0.23$\pm$0.30 & --- \\
% coma: g-r=0.421+/-0.019; r-i=0.079+/-0.011
2021 Jul 13 & FTN & & 0.47$\pm$0.02 & 0.13$\pm$0.03 & 0.05$\pm$0.04 & & 0.45$\pm$0.30 & 0.16$\pm$0.30 & $-$0.01$\pm$0.40 \\
% coma: g-r=0.507+/-0.021; r-i=0.132+/-0.025; i-z=0.060+/-0.040
2021 Jul 15 & FTN &  & 0.50$\pm$0.02 & 0.10$\pm$0.03 & 0.03$\pm$0.04 & & 0.51$\pm$0.30 & 0.19$\pm$0.30 & 0.01$\pm$0.40 \\
% coma: g-r=0.502+/-0.020; r-i=0.098+/-0.025; i-z=0.028+/-0.034
\multicolumn{1}{c}{\it Median$^a$} & ... & & 0.47$\pm$0.03 & 0.10$\pm$0.04 & 0.05$\pm$0.05 & & 0.51$\pm$0.04 & 0.19$\pm$0.07 & $-$0.01$\pm$0.01 \\
% 2021-07-09_Palomar_coma: g-r = 0.471+/-0.009; r-i = 0.134+/-0.007
% 2021-07-09_Palomar_tail: g-r = 0.470+/-0.234; r-i = 0.025+/-0.255
% 2021-07-09_FTN_coma: g-r = 0.535+/-0.019; r-i = 0.034+/-0.027; i-z = 0.110+/-0.040
% 2021-07-09_FTN_tail: g-r = 0.574+/-0.183; r-i = 0.169+/-0.243; i-z = 0.001+/-0.376
% 2021-07-12_Palomar_coma: g-r = 0.421+/-0.019; r-i = 0.079+/-0.011
% 2021-07-12_Palomar_tail: g-r = 0.518+/-0.292; r-i = 0.252+/-0.127
% 2021-07-13_FTN_coma: g-r = 0.507+/-0.021; r-i = 0.132+/-0.025; i-z = 0.060+/-0.040
% 2021-07-13_FTN_tail: g-r = 0.451+/-0.287; r-i = 0.123+/-0.348; i-z = -0.054+/-0.555
% 2021-07-15_FTN_coma: g-r = 0.502+/-0.020; r-i = 0.098+/-0.025; i-z = 0.028+/-0.034
% 2021-07-15_FTN_tail: g-r = 0.557+/-0.228; r-i = 0.233+/-0.279; i-z = -0.047+/-0.338
% Mean_coma: g-r = 0.4811+/-0.0066; r-i = 0.1155+/-0.0055; i-z = 0.0633+/-0.0218
% Mean_tail: g-r = 0.5292+/-0.1048; r-i = 0.2062+/-0.0933; i-z = -0.0296+/-0.2290
\hline
\hline
\multicolumn{10}{l}{$^a$ Median values of computed colors, where standard deviations are used as uncertainties.} \\
\end{tabular}
\label{table:obs_2005qn173_colors}
\end{table*}

\section{Results and Analysis}\label{section:results}

\subsection{Data Analysis\label{section:data_analysis}}

Except for data from 2016 July 22, 248370 had a star-like surface brightness profile in all archival images and exhibited no other visible indications of activity. Meanwhile, in all 2021 observations, the object exhibited a long, straight dust tail oriented along the coincident antisolar and negative heliocentric velocity vector directions as projected on the sky.  In our best composite image from UT 2021 July 12, the tail was seen extending $\sim9'$ from the nucleus (Figure~\ref{figure:images}f), corresponding to a physical extent of $\sim$720,000 km at the geocentric distance of the comet.  Minimal coma was present in all images, with full-width at half-maximum (FWHM) measurements of the nucleus's surface brightness profile measured in the direction perpendicular to the dust tail nearly identical to FWHM measurements, $\theta_s$, of field star profiles (listed in Table~\ref{table:obs_2005qn173_active}), measured in the direction perpendicular to their trailing due to non-sidereal tracking.  We did however find the half-width at half-maximum (HWHM) of the object's profile measured along the sunward direction directly opposite the dust tail to be $\sim$10\% larger than stellar HWHM values.

To maximize signal-to-noise ratios (S/N) for sets of observations where more than one image was obtained in the same filter in a night, we constructed composite images by shifting and aligning individual images in each filter on the object's photocenter using linear interpolation, and adding them together.  Representative single or composite images are shown in Figure~\ref{figure:images}.

For photometry of all data, measurements of 248370 and 10-30 nearby reference stars were performed using
%Image Reduction and Analysis Facility 
IRAF software \citep[][]{tody1986_iraf,tody1993_iraf}, with absolute calibration performed using field star magnitudes in Sloan bandpasses derived from the {\tt RefCat2} all-sky catalog \citep[which uses the PS1 photometric system;][]{tonry2018_refcat}.  Nucleus or near-nucleus coma photometry of 248370 was performed using circular apertures with sizes chosen using curve-of-growth analyses when the object appeared inactive, or circular apertures with fixed radii equivalent to 5000~km at the geocentric distance of the object when it was active.  For the latter, photometry aperture radii, $\theta_{\rm obs}$, were determined from convolving the projected angular equivalent, $\theta_0$, of the desired intrinsic distance (i.e., 5000~km) at the geocentric distance of the comet with the FWHM seeing, $\theta_s$, on a given night using
\begin{equation}
    \theta_{\rm obs} = \left(\theta_0^2 + \theta_s^2 \right)^{1/2}
    \label{equation:convolve}
\end{equation}
where $\theta_{\rm obs}\sim4''$ for most of our observations.
%For reference, the coma had a FWHM of $1\farcs4$ to $2\farcs4$ (e.g., see Section~\ref{section:dust_ejection}).
Background statistics for comet photometry were measured in nearby regions of blank sky to avoid dust contamination from the object or nearby field stars.

%For archival data where the object was stellar in appearance in all cases, optimal radii for photometry apertures were selected based on curve of growth analyses.  Meanwhile, for new data where the object exhibited visible coma, we used photometry apertures with radii of $4''$ in all cases to characterize the combined properties of the nucleus and near-nucleus coma.

We also measured the surface brightnesses of 248370's dust tail on each night by rotating composite images to make the dust tail horizontal in each image frame, measuring net fluxes in rectangular apertures placed along the length of each tail, and converting those fluxes to surface brightnesses in mag~arcsec$^{-2}$ using the measured mean magnitudes of the nucleus for data comprising each composite image for absolute photometric calibration.  We chose rectangular apertures that extended 750~km above and below the tail's central axis in the vertical direction (i.e., $\sim2''$ in total height for most of our observations) and from 5000~km to 15000~km (i.e., from $\sim4''$ to $\sim12''$ for most of our observations) from the nucleus in the horizontal direction, where the angular sizes of these apertures were computed in the same manner described earlier for near-nucleus photometry apertures. The details of this method of measuring surface brightnesses were chosen to maximize signal-to-noise while also minimizing nucleus flux contribution by focusing on the bright central core of the tail, and measuring close, but not too close, to the nucleus where the tail is brightest.

Photometric results for data obtained when 248370 appeared active and inactive are shown in Tables~\ref{table:obs_2005qn173_active} and \ref{table:obs_2005qn173_inactive}, respectively.  Colors computed by comparing coma magnitudes or tail surface brightnesses in different filters for nights on which multi-filter data were obtained are shown in Table~\ref{table:obs_2005qn173_colors}.

\subsection{Nucleus Properties\label{section:nucleus}}

Using measured apparent magnitudes of 248370 from archival data, we derive magnitudes normalized to $r_h=\Delta=1$~au and $\alpha=0^{\circ}$, or $m(1,1,0)$, by assuming inverse-square-law fading and IAU $H,G$ phase function behavior \citep{bowell1989_astphotmodels_ast2} where $G=0.15$ (Table~\ref{table:obs_2005qn173_inactive}). We then take medians of these computed $m(1,1,0)$ values to estimate absolute magnitudes in each filter.
We estimate 248370's absolute magnitudes to be $H_g=16.62\pm0.13$, $H_r=16.12\pm0.10$, $H_i=16.05\pm0.11$, and $H_z=15.93\pm0.08$ (Table~\ref{table:obs_2005qn173_inactive}), corresponding to nucleus colors of $g'-r'=0.50\pm0.16$, $r'-i'=0.07\pm0.15$, and $i'-z'=0.12\pm0.14$, which within uncertainties, are effectively solar \citep[e.g.,][]{holmberg2006_solarcolors}.  These colors are consistent with a C-type taxonomic classification \citep[see][]{demeo2013_asteroidtaxonomy}, which is the most likely classification expected for an outer main belt asteroid like 248370, but large uncertainties on the colors derived here from sparse archival data mean that other taxonomic types cannot necessarily be excluded at this time.  Using $V=g'-0.565(g'-r') - 0.016$ \citep{jordi2006_filtertransformations}, we find an equivalent $V$-band absolute magnitude of $H_V=16.32\pm0.10$.

Using
\begin{equation}
r_n = \left( {2.24\times10^{22}\over p_V} \times 10^{0.4(m_{\odot,V} - H_V)} \right)^{1/2}
\end{equation}
where we use $p_V=0.054\pm0.012$ \citep{mainzer2019_neowisealbedos_v2} for the object's $V$-band albedo and $m_{\odot,V}=-26.71\pm0.03$ for the apparent $V$-band magnitude of the Sun \citep{hardorp1980_sun3}, we find an effective nucleus radius of $r_n=1.6\pm0.2$~km, or slightly smaller than the radius computed by \citet{mainzer2019_neowisealbedos_v2}.

The ranges in computed absolute magnitudes
%\comment{the object's variability instead of the ranges in computed absolute magnitudes}
in each filter are $\Delta m_g=0.29$~mag, $\Delta m_r=0.25$~mag, $\Delta m_i=0.23$~mag, and $\Delta m_z=0.13$.  These values are not particularly meaningful given the small number of data points used to derive them, but in the present absence of better measurements, they suggest that 248370's photometric range due to rotation is $\Delta m\gtrsim0.3$~mag, implying a minimum axis ratio of $a/b=1.3$.

\subsection{Activity Properties\label{section:activity}}

\subsubsection{Dust Composition\label{section:dust_composition}}

%Using measured near-nucleus apparent magnitudes of 248370 while it was active in 2021, we are able to characterize various aspects of its activity.
We find mean coma colors of $g'-r'=0.47\pm0.03$, $r'-i'=0.10\pm0.04$, and $i'-z'=0.05\pm0.05$ and mean dust tail colors of $g'-r'=0.51\pm0.04$, $r'-i'=0.19\pm0.07$, and $i'-z'=-0.01\pm0.01$ (Table~\ref{table:obs_2005qn173_colors}).
Within uncertainties, coma and dust tail colors are comparable to one another, indicating that both are dominated by dust of similar composition with no apparent color gradient with distance from the nucleus that might indicate the presence of a significant near-nucleus gas coma.
The apparent compositional similarity of coma and tail dust also means that we see no evidence of grain fragmentation or loss of icy grains to sublimation that could cause overall color changes to observed dust.
%you could also say that there is no evidence of the grains changing color, which might indicate grain fragmentation or loss of icy grains
The colors of both are also similar within uncertainties to colors found for the bare nucleus (Section~\ref{section:nucleus}), suggesting that the dust coma and tail are compositionally similar to the nucleus's surface regolith.

\subsubsection{Activity Strength and Evolution\label{section:coma_evolution}}

From our calculations of 248370's absolute magnitudes (Section~\ref{section:nucleus}), we find that the near-nucleus region of the object was $\sim$0.5~mag brighter than expected for the inactive nucleus on 2016 July 22 and $\sim$1~mag brighter than expected in 2021  (Table~\ref{table:obs_2005qn173_active}).
We also compute the ratios, $A_d/A_n$, of the scattering cross-sections of ejected near-nucleus dust within our 5000~km photometry apertures and the underlying nucleus when 248370 was active using
\begin{equation}
    A_d/A_n = {{1 - 10^{0.4(m(1,1,0)-H)}}\over{10^{0.4(m(1,1,0)-H)}}}
\end{equation}
\citep[e.g.,][]{hsieh2021_259p}.
We find $A_d/A_n=0.7\pm0.3$ on 2016 July 22, and $1.8<A_d/A_n<2.9$ in 2021 (Table~\ref{table:obs_2005qn173_active}).

\begin{figure*}[tbp]
\centerline{\includegraphics[width=4.0in]{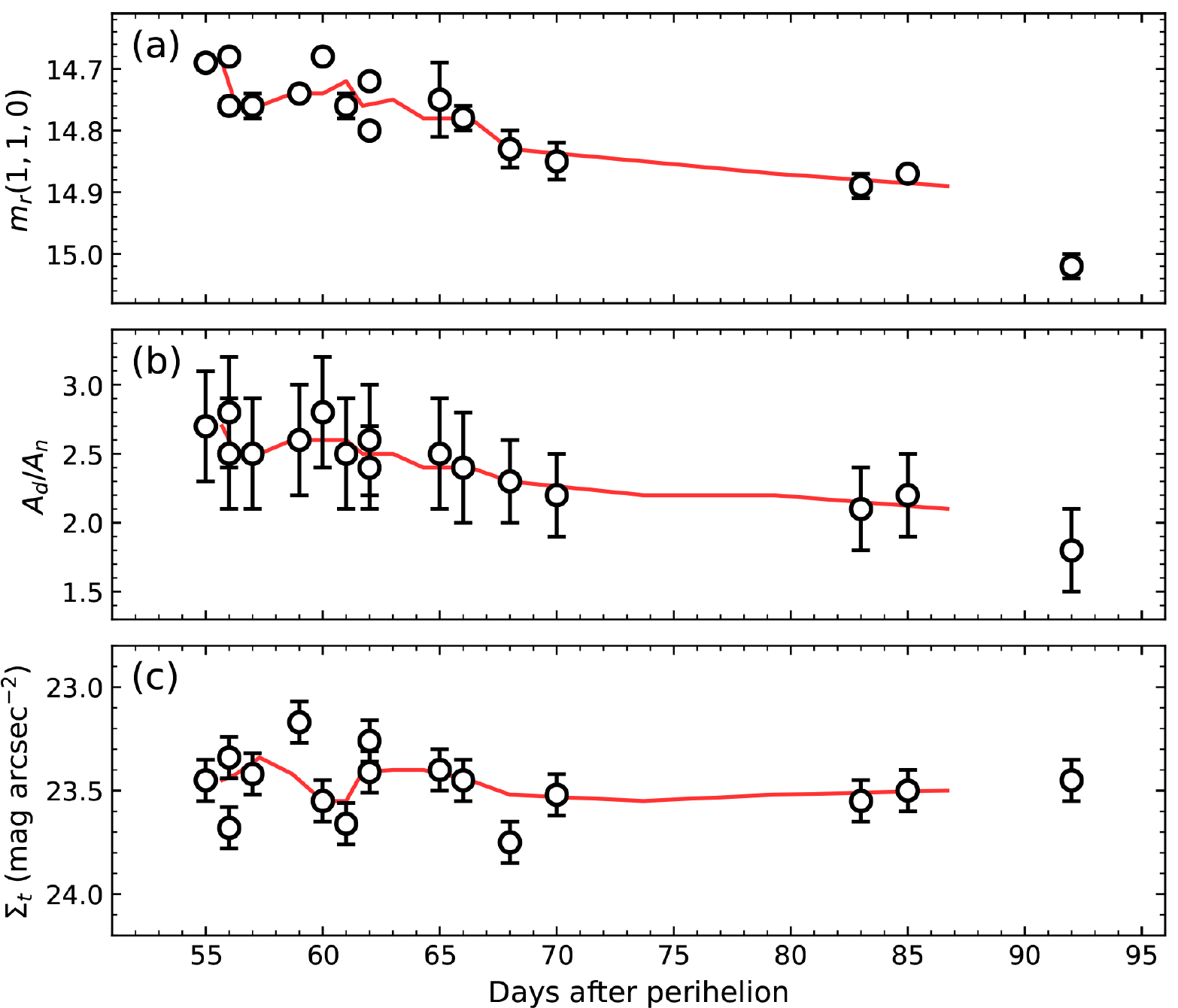}}
\caption{\small Plots of (a) $r'$-band near-nucleus magnitudes normalized to $r_h=\Delta=1$~au and $\alpha=0^{\circ}$, (b) inferred ratios of dust scattering cross-sections to nucleus scattering cross-sections measured within 5000~km-radius photometry apertures, and (c) surface brightnesses in mag~arcsec$^{-1}$ of a fixed portion of the dust tail, computed for 248370 from UT 2021 July 8 to 2021 August 14 as a function of days after perihelion (UT 2021 May 14), where red lines in each panel show moving medians (computed for groups of three data points each) for each quantity.}
\label{figure:fading}
\end{figure*}

Plotting $m(1,1,0)$ and $A_d/A_n$ as functions of time, we see that the coma faded during our 2021 observations (Figures~\ref{figure:fading}a and \ref{figure:fading}b), declining in intrinsic brightness by $\sim$0.35 mag (or $\sim$25\%) in 37 days.  Increasing activity strength would suggest ongoing dust production, and therefore the action of a prolonged, possibly sublimation-driven, emission event.  However, declining activity strength does not necessarily rule out a sublimation-driven emission event, especially at the relatively gradual rate ($\sim$0.01~mag/day) seen for 248370, similar to the rate of fading of the coma of confirmed recurrently active MBC 259P/Garradd
\citep{hsieh2021_259p} of $\sim$0.015~mag/day observed after its discovery in 2008 \citep{jewitt2009_259p}.

Despite the fading of 248370's coma, the dust tail remained relatively consistent in brightness during our observations (Table~\ref{table:obs_2005qn173_active}; Figure~\ref{figure:fading}c), suggesting that the tail may consist of larger particles on average than the coma. Larger particles in the tail would be dissipated by radiation pressure more slowly than presumably smaller particles in the coma, which would explain the slower fading of the tail to apparently weakening dust production from the nucleus.

For reference, we also compute $A(\alpha=0^{\circ})f\rho$ values (hereafter, $Af\rho$), which are nominally independent of photometry aperture sizes for observations of comae with $r^{-1}$ radial profiles, and are given by
\begin{equation}
A(\alpha=0^{\circ})f\rho = {(2r_h\Delta)^2\over \rho}10^{0.4[m_\odot-m_{d}(r_h,\Delta,0)]}
\label{equation:afrho}
\end{equation}
\citep{ahearn1984_bowell}, where $r_h$ is in au, $\Delta$ is in cm, $\rho$ is the physical radius in cm of the photometry aperture at the geocentric distance of the comet, $m_{\odot}$ is the Sun's apparent magnitude in the specified filter \citep[using $m_{g,\odot}=-26.60$, $m_{r,\odot}=-27.05$, $m_{i,\odot}=-27.17$, and $m_{z,\odot}=-27.21$;][]{hardorp1980_sun3,jordi2006_filtertransformations,holmberg2006_solarcolors}, and $m_{d}(r_h,\Delta,0)$ is the phase-angle-normalized (to $\alpha=0^{\circ}$) magnitude of the excess dust mass of the comet (i.e., with the flux contribution of the nucleus subtracted out).  These results are tabulated in Table~\ref{table:obs_2005qn173_active},
%and plotted in Figure~\ref{figure:fading}c, 
where we see fading behavior similar to that seen for $m(1,1,0)$ and $A_d/An$.

%\msk{Delta is changing, so a 4 arcsec aperture corresponds to 1390 km at the start, but 1280 km at the end of our photometry, i.e., the effective size at the distance of the comet *shrinks* by 9\%, which will decrease the amount of dust within the aperture.  The Afrho model probably isn't accounting for the difference because it assumes a 1/rho surface brightness distribution.  The fact that Afrho is also decreasing could mean the activity is fading, but couldn't it also be interpreted as that the mean surface brightness profile is shallower than 1/rho?  Or, a combination of the two scenarios?  What are the results for a fixed-linear aperture radius, e.g., picking a nice round number, 1400 km?}
%We note, however, that this parameter is not always a reliable measurement of the dust contribution to comet photometry in cases of non-spherically symmetric comae \citep[e.g.,][]{fink2012_afrho}.

\subsubsection{Dust Ejection Parameter Constraints\label{section:dust_ejection}}

Dust grains ejected sunward with a terminal ejection velocity of $v_g$ are turned back by solar radiation pressure on a distance scale given by
\begin{equation}
    X_R \sim {v_g^2r_h[{\rm au}]^2\over2\beta_d g_{\odot}}
    \label{equation:turnback}
\end{equation}
\citep{jewitt1987_cometsbps}, where $r_h[{\rm au}]$ is the heliocentric distance in au, $g_{\odot}=0.006$~m~s$^{-2}$ is the gravitational acceleration to the Sun at 1~au, %import astropy.constants as const
%print(const.GM\_sun / const.au**2) = 0.0059300835181528 m / s2} ,
and $\beta_d$ is the ratio of the acceleration experienced by a particle due to solar radiation pressure to the local acceleration due to solar gravity \citep{burns1979_dustradiationforces}.  Comet dust modeling analyses commonly use $\beta_d$ to represent particle sizes, where $a_d\approx\beta_d^{-1}$ gives approximate corresponding dust particle radii, $a_d$, in $\mu$m.  

On UT 2021 July 12, we measure a HWHM value for the sunward portion of the coma of $\theta_{\rm obs}/2=0\farcs68$, while nearby field stars had HWHM values of $\theta_s/2=0\farcs61$.  Using an analogous form of Equation~\ref{equation:convolve} to compute an inferred intrinsic half-width of the coma, $\theta_0/2$, in the absence of atmospheric seeing, we find $\theta_0/2=0\farcs3$, or $\sim$400~km at the geocentric distance of the object.  Using $X_R=400$~km, Equation~\ref{equation:turnback} gives $v_g\sim0.9\beta_d^{1/2}$~m~s$^{-1}$, or about half the ejection velocities found for 133P/Elst-Pizarro \citep{jewitt2014_133p}, another MBC very similar in morphology to 248370.

Analyzing the composite image for UT 2021 July 12, we measure a median FWHM of $\theta_{\rm obs}=1\farcs47$
%(with a standard deviation of $\sigma=0.16$)
for the dust tail over the $2'$ of the tail closest to the nucleus, as measured in intervals of 20 pixels ($3\farcs5$) along the tail.  The tail's FWHM flares slightly from $\theta_{\rm obs}=1\farcs45$ close to the nucleus to $\theta_{\rm obs}=1\farcs50$ at $2'$ from the nucleus.  Using Equation~\ref{equation:convolve}, we find an intrinsic median tail FWHM of $\theta_0=0\farcs8\pm0\farcs3$, corresponding to a physical projected width of $\sim$1100~km in the plane of the sky.  The narrowness of the tail is consistent with low dust ejection velocities, and similar to that found for 133P \citep{hsieh2004_133p}.

Performing a simple (zero-ejection velocity) dust modeling analysis using the online Comet Toolbox\footnote{\url{http://comet-toolbox.com/FP.html}}, we find that particles with $\beta=1$ (or $a\sim1~\mu$m)
%typically the largest dust particles that are considered in comet dust modeling analyses; e.g.,][]{ishiguro2007_spcs} 
would take $\sim$20~days to reach an apparent angular separation from the nucleus of $\sim$9$'$ (the visible length of the tail on UT 2021 July 12; Figure~\ref{figure:images}f).  Meanwhile, particles with $\beta=0.1$, $\beta=0.01$, and $\beta=0.001$ (or $a\sim10$~$\mu$m, $a\sim100$~$\mu$m, and $a\sim1$~mm), which span the range of particle sizes found for other MBCs \citep[e.g.,][]{hsieh2009_238p,licandro2013_288p,jewitt2014_133p}, would take 60, 150, and 430 days, respectively, to reach the same apparent separation.  Without additional particle size constraints at the present time, however, we are unable to meaningfully constrain the likely ejection times of the most distant dust grains in 248370's tail.
%from the length of the tail alone.
%, except that they were likely ejected no later than 20 days prior to 2021 July 12 (assuming $\beta_{\rm max}=1$, or 2021 June 22 when the object was at $r_h=2.383$~au and $\nu=11.6^{\circ}$, but could have been (and most likely were) ejected much longer ago.
We note that if activity began when 248370 was at $\nu=300^{\circ}$ \citep[the earliest activation point confirmed to date for a MBC;][]{hsieh2015_324p}, which the object passed on 2020 October 22, particles as large as $a\sim400$~$\mu$m ($\beta=0.0025$) would have been able to reach a $9'$ separation from the nucleus by 2021 July 12.

%\subsection{Comparison to Other Active Asteroids\label{section:other_AAs}}

%With a radius of $r_n=1.6\pm0.2$~km, 248370 is one of the larger MBC nuclei to be measured, smaller than just 133P and 176P \citep[$r_n=1.9$~km and $r_n=2.0$~km, respectively;][and references within]{hsieh2009_albedos,snodgrass2017_mbcs}.  Like 133P

%\hhh{compare nucleus sizes, peak of activity, }

%248370 bears a striking similarity to the first MBC to be discovered, 133P/Elst-Pizarro.  

\subsection{Future Work\label{section:futurework}}

The discovery that 248370 is recurrently active near perihelion strongly suggests that sublimation is a primary driver of its activity, although it does not rule out other processes that could also contribute to the current observed activity or may have triggered it.  In particular, the slow terminal velocities inferred for ejected dust grains (Section~\ref{section:dust_ejection}) suggest that rapid rotation, nucleus elongation, or both could be acting to reduce or negate the effective gravity felt by dust particles at certain locations on the nucleus surface, allowing even extremely large particles to be ejected, similar to what may be occurring on 133P \citep{jewitt2014_133p}.
As such, measurement of 248370's rotational period and nucleus shape, as well as its taxonomic type, should be considered a high priority once its current activity ends.  Continued monitoring of 248370's current activity is also highly encouraged to enable further characterization of the object's fading behavior, which can help constrain the dust grain size distribution.

%Using expressions for the amount of damping of lightcurve amplitudes by a coma with a given cross-section ratio with respect to an underlying bare nucleus,

A detailed dynamical analysis of 248370 is outside the scope of this paper, but should also be performed in the near future.  Issues to consider include whether the object can be linked to any dynamical asteroid families \citep[e.g.,][]{hsieh2018_activeastfamilies}, its long-term dynamical stability and whether it may be an implanted object \citep[e.g.,][]{hsieh2016_tisserand}, and whether it follows dynamical trends found for previously discovered MBCs \citep{kim2018_mbcalignment}.

In the long term, 248370 will be well-placed for monitoring during the approach to its next perihelion passage on UT 2026 September 3.  It becomes observable from the Southern Hemisphere in 2026 February at $\nu\sim300^{\circ}$, i.e., the earliest activation point confirmed to date for a MBC, as discussed earlier.  Monitoring during this time will be extremely valuable for further confirming the recurrent nature of 248370's activity, constraining the orbital range over which activity occurs (with implications for constraining ice depth on the object, as well as its active lifetime), measuring initial dust production rates, and comparing the object's activity levels from one orbit to another as well as to other MBCs.

\clearpage

\acknowledgments

HHH, COC, MSPK, MMK, JP, SSS, AT, and CAT acknowledge support from the NASA Solar System Observations program (Grant 80NSSC19K0869).  COC is also supported by the National Science Foundation Graduate Research Fellowship Program under Grant No.\ (2018258765). Any opinions, findings, and conclusions or recommendations expressed in this material are those of the author(s) and do not necessarily reflect the views of the National Science Foundation.  The work of JP was conducted at the Jet Propulsion Laboratory, California Institute of Technology, under a contract with the National Aeronautics and Space Administration (80NM0018D0004).
HU, TA, RM, PR, and BW were supported by a UK Science and Technology Facilities Council (STFC) 'Spark Award' grant for the Comet Chasers school outreach program.

We are grateful to K.\ Peffer, P.\ Nied, C.\ Heffner, and T.\ Barlow for assistance in obtaining observations with Palomar, and I.\ Nisley and C.\ Siqueiros for assistance in obtaining observations with LDT.
This research made use of {\tt astropy}, a community-developed core {\tt python} package for astronomy, {\tt uncertainties} (version 3.0.2), a {\tt python} package for calculations with uncertainties by E.\ O.\ Lebigot, scientific software at {{\url{http://www.comet-toolbox.com}}} \citep{vincent2014_comettoolbox}, and NASA's Astrophysics Data System Bibliographic Services.

%This work benefited from support by the International Space Science Institute in Bern, Switzerland, through the hosting and provision of financial support for an international team, which was led by C.\ Snodgrass and included HHH and MMK, to discuss the science of MBCs.  
%This work is based on observations obtained at the Gemini Observatory, which is operated by the Association of Universities for Research in Astronomy, Inc., under a cooperative agreement with the NSF on behalf of the Gemini partnership: the National Science Foundation (United States), the National Research Council (Canada), CONICYT (Chile), Ministerio de Ciencia, Tecnolog\'{i}a e Innovaci\'{o}n Productiva (Argentina), and Minist\'{e}rio da Ci\^{e}ncia, Tecnologia e Inova\c{c}\~{a}o (Brazil).

This research made use of data obtained at the Lowell Discovery Telescope (LDT).  Lowell Observatory is a private, nonprofit institution dedicated to astrophysical research and public appreciation of astronomy and operates the LDT in partnership with Boston University, the University of Maryland, the University of Toledo, Northern Arizona University, and Yale University. Partial support of the LDT was provided by Discovery Communications. LMI was built by Lowell Observatory using funds from the National Science Foundation (NSF grant AST-1005313), PI: P.\ Massey).
This work also used observations obtained at the Hale Telescope at Palomar Observatory, which is operated as part of a continuing collaboration between the California Institute of Technology, NASA/JPL, Oxford University, Yale University, and the National Astronomical Observatories of China.

This work also includes observations from the Las Cumbres Observatory global telescope network.  Specifically, this paper uses observations made with the MuSCAT3 instrument, developed by the Astrobiology Center and with financial support from JSPS KAKENHI (JP18H05439) and JST PRESTO (JPMJPR1775), at Faulkes Telescope North on Maui, HI, operated by the Las Cumbres Observatory.

Observations with the LCOGT 1m were obtained as part of the LCOGT Outbursting Objects Key (LOOK) Project (KEY2020B-009), while FTN observations were obtained via the aforementioned Comet Chasers school outreach program. The Comet Chasers program is part of the Faulkes Telescope Project (FTPEPO2014A-004), which is partly funded by the Dill Faulkes Educational Trust.  Pupils from three schools in Wales (St Mary's Catholic Primary School Bridgend, Mount Street Junior School Brecon, and Ynysowen Community Primary School) made observations.

The Asteroid Terrestrial-impact Last Alert System (ATLAS) project is primarily funded to search for near earth asteroids through NASA grants NN12AR55G, 80NSSC18K0284, and 80NSSC18K1575. The ATLAS science products have been made possible through the contributions of the University of Hawaii Institute for Astronomy, the Queen’s University Belfast, the Space Telescope Science Institute, the South African Astronomical Observatory, and The Millennium Institute of Astrophysics (MAS), Chile.

This research used the facilities of the Canadian Astronomy Data Centre operated by the National Research Council of Canada with the support of the Canadian Space Agency.
The Pan-STARRS1 Surveys (PS1) and the PS1 public science archive have been made possible through contributions by the Institute for Astronomy, the University of Hawaii, the Pan-STARRS Project Office, the Max-Planck Society and its participating institutes, the Max Planck Institute for Astronomy, Heidelberg and the Max Planck Institute for Extraterrestrial Physics, Garching, The Johns Hopkins University, Durham University, the University of Edinburgh, the Queen's University Belfast, the Harvard-Smithsonian Center for Astrophysics, the Las Cumbres Observatory Global Telescope Network Incorporated, the National Central University of Taiwan, the Space Telescope Science Institute, the National Aeronautics and Space Administration (NASA) under Grant NNX08AR22G, the National Science Foundation (NSF) under Grant AST-1238877, the University of Maryland, Eotvos Lorand University (ELTE), the Los Alamos National Laboratory, and the Gordon and Betty Moore Foundation.

This project used data obtained with the Dark Energy Camera (DECam), which was constructed by the Dark Energy Survey (DES) collaborating institutions: Argonne National Lab, University of California Santa Cruz, University of Cambridge, Centro de Investigaciones Energeticas, Medioambientales y Tecnologicas-Madrid, University of Chicago, University College London, DES-Brazil consortium, University of Edinburgh, ETH-Zurich, University of Illinois at Urbana-Champaign, Institut de Ciencies de l'Espai, Institut de Fisica d'Altes Energies, Lawrence Berkeley National Lab, Ludwig-Maximilians Universitat, University of Michigan, National Optical Astronomy Observatory, University of Nottingham, Ohio State University, University of Pennsylvania, University of Portsmouth, SLAC National Lab, Stanford University, University of Sussex, and Texas A\&M University. Funding for DES, including DECam, has been provided by the U.S.\ Department of Energy, National Science Foundation, Ministry of Education and Science (Spain), Science and Technology Facilities Council (UK), Higher Education Funding Council (England), National Center for Supercomputing Applications, Kavli Institute for Cosmological Physics, Financiadora de Estudos e Projetos, Funda{\c c}{\=a}o Carlos Chagas Filho de Amparo a Pesquisa, Conselho Nacional de Desenvolvimento Cient{\'i}fico e Tecnol{\'o}gico and the Minist{\'e}rio da Ci{\^e}ncia e Tecnologia (Brazil), the German Research Foundation-sponsored cluster of excellence ``Origin and Structure of the Universe'' and the DES collaborating institutions.

The national facility capability for SkyMapper has been funded through ARC LIEF grant LE130100104 from the Australian Research Council, awarded to the University of Sydney, the Australian National University, Swinburne University of Technology, the University of Queensland, the University of Western Australia, the University of Melbourne, Curtin University of Technology, Monash University and the Australian Astronomical Observatory. SkyMapper is owned and operated by The Australian National University's Research School of Astronomy and Astrophysics. The survey data were processed and provided by the SkyMapper Team at ANU. The SkyMapper node of the All-Sky Virtual Observatory (ASVO) is hosted at the National Computational Infrastructure (NCI). Development and support of the SkyMapper node of the ASVO has been funded in part by Astronomy Australia Limited (AAL) and the Australian Government through the Commonwealth's Education Investment Fund (EIF) and National Collaborative Research Infrastructure Strategy (NCRIS), particularly the National eResearch Collaboration Tools and Resources (NeCTAR) and the Australian National Data Service Projects (ANDS).

Finally, this work made use of observations obtained with MegaPrime/MegaCam, a joint project of CFHT and CEA/DAPNIA, at the Canada-France-Hawaii Telescope (CFHT) which is operated by the National Research Council (NRC) of Canada, the Institut National des Science de l'Univers of the Centre National de la Recherche Scientifique (CNRS) of France, and the University of Hawaii. The observations at the Canada-France-Hawaii Telescope were performed with care and respect from the summit of Maunakea which is a significant cultural and historic site.

%% To help institutions obtain information on the effectiveness of their 
%% telescopes the AAS Journals has created a group of keywords for telescope 
%% facilities.
%
%% Following the acknowledgments section, use the following syntax and the
%% \facility{} or \facilities{} macros to list the keywords of facilities used 
%% in the research for the paper.  Each keyword is check against the master 
%% list during copy editing.  Individual instruments can be provided in 
%% parentheses, after the keyword, but they are not verified.

\vspace{5mm}
\facilities{Blanco (DECam), CFHT (MegaCam), PS1, LDT (LMI), Hale (WASP), LCOGT, FTN (MuSCAT3), Skymapper}

%% Similar to \facility{}, there is the optional \software command to allow 
%% authors a place to specify which programs were used during the creation of 
%% the manuscript. Authors should list each code and include either a
%% citation or url to the code inside ()s when available.

\software{{\tt astropy} \citep{astropy2018_astropy},
    {\tt astroquery} \citep{ginsburg2019_astroquery},
    {\tt IRAF} \citep{tody1986_iraf,tody1993_iraf}, 
    {\tt L.A.Cosmic} \citep{vandokkum2001_lacosmic,vandokkum2012_lacosmic},
    {\tt uncertainties} (v3.0.2, E.\ O.\ Lebigot),
    {\tt RefCat2} \citep{tonry2018_refcat},
    Comet-Toolbox \citep{vincent2014_comettoolbox}
          }

%% Appendix material should be preceded with a single \appendix command.
%% There should be a \section command for each appendix. Mark appendix
%% subsections with the same markup you use in the main body of the paper.

%% Each Appendix (indicated with \section) will be lettered A, B, C, etc.
%% The equation counter will reset when it encounters the \appendix
%% command and will number appendix equations (A1), (A2), etc. The
%% Figure and Table counter will not reset.

%% For this sample we use BibTeX plus aasjournals.bst to generate the
%% the bibliography. The sample63.bib file was populated from ADS. To
%% get the citations to show in the compiled file do the following:
%%
%% pdflatex sample63.tex
%% bibtext sample63
%% pdflatex sample63.tex
%% pdflatex sample63.tex

\clearpage

\bibliography{main}{}
\bibliographystyle{aasjournal}

\end{document}